# Three-plate graphene capacitor for high-density electric energy storage

Alexander Khitun

*Department of Electrical and Computer Engineering, University of California - Riverside, Riverside, California, USA 92521*

Correspondence to akhitun@engr.ucr.edu

**Abstract.** Graphene possesses a unique combination of physical properties including high carrier mobility and high current density it can sustain. In contrast to bulk metals, graphene does not completely screen the external electrostatic field. In this work, we consider the possibility of utilizing these properties for building devices for high-density electric energy storage. We consider a three-plate parallel plate capacitor where the middle plate is made of graphene and negatively charged. The electric forces of attraction acting on the electrons on the middle plate are compensated as the electric fields on both sides of the plate are not screened. However, it brings the system to an unstable equilibrium state. To make system stable, we consider fast oscillations similar to ones in Kapitza's pendulum. AC electric current through the middle graphene plane creates a magnetic field. In turn, Lorentz force squeezes moving electrons towards the center of the middle plane. We present the results of numerical modeling showing the effect of AC electric current on electron movement. According to the estimates, the pseudopotential produced by the AC current may exceed $60\ eV$ at room temperature. Such a large value of the pseudopotential is due to the high mobility and large current density in graphene. The electric field intensity between the edge plates and the middle plate may exceed the breakdown value for a conventional double-plate parallel-plate capacitors leading to the increase in the electric energy storage. The electric breakdown of the graphene capacitor is limited by the mechanical strength of the side plates. It may be possible to enhance the volume electric energy density above the gasoline 34 MJ/L. We also describe possible experiments to validate this idea.

The discovery of graphene in 2004 opened a new field of research aimed at exploring 2D materials [1]. Graphene has appeared to have a unique combination of physical properties including high carrier mobility [2], high thermal conductivity [3], and high mechanical strength[4] which translated in a variety of practical applications [5-11]. During the past two decades, there were more than 300,000 scientific publications devoted to graphene [12]. Still yet, graphene keeps its mystery to be further explored. There still remains a simple unanswered question regarding the way in which the induced charge density is distributed in the vicinity of a graphene monolayer when an external electric field is applied to the graphene sheet [13]. The perpendicular charge distribution was studied by considering two- and multilayer graphene films [14,15]. The charge distribution in a monolayer graphene has been explored less. It is estimated to be within the angstrom-scale distances [13]. In contrast to bulk metals, 2D conductors like graphene do not completely screen the external static electric field [16]. Here, we consider the possibility of exploiting this graphene property for energy density enhancement in electrostatic capacitors.

To explain the idea, we compare two three-plate capacitors. In one case, the middle plate is made of a bulk metal and in the second case, the middle plate is made of graphene. In Fig.1(a), there is shown a three-plate parallel plate capacitor where the central plate is made of a regular (bulk) metal whose thickness is much larger than the inter-atomic distance. The plates are shown in the X-Z plane. The area of the plate is S. The plates are separated by distance d. The space between the plates is filled with a vacuum. The side plates are positively charged storing charge +q each. The negative charge of -2q is stored on the middle plate. The charge on the middle plate is accumulated on the sides of the plate. The electric field is non-zero between the plates and vanishes inside the metallic plate. The electron potential energy $U$ along the Y-axis is depicted in Fig.1(b). Starting from the left plate, it linearly increases and reaches its maximum on the middle plate. It keeps constant on the middle plate and decreases linearly from the edge of the middle plate to the plate on the right. The derivative of the potential is the force acting on the electrons. There is a force of attraction (i.e., Coulomb force) acting on the charges on the sides of the middle plate. The force is uncompensated as the internal electric field inside the metallic plate is totally screened. The maximum amount of energy that can be stored in this capacitor is limited by the critical value of the electric breakdown field $E_b$.

Next, let us consider a three-plate capacitor where the middle plate is made of graphene as shown in Fig.1(c). We assume no charge separation within the atomically thin graphene layer. In Fig.1(d), it is shown the profile of electron potential energy for the graphene-based three-plate capacitor. The electron potential energy reaches its peak at the graphene layer. There is a net zero force acting on the electrons on the middle plate. The forces of attraction acting on the charge compensate each other in the center of the middle plate. This is possible if and only if the electric field penetrates graphene. In this case, the maximum electric field between the middle and the side plates is limited only by the mechanical strength of the side plates (i.e., the electric field at which the positively charged ions of the plates will be emitted from the side plates) and the dielectric strength of the vacuum. However, the negative charges on the middle plate are in the unstable equilibrium. A small deviation from the central position (e.g., caused by a thermal fluctuation) may destroy the equilibrium and lead to an electric breakdown if the electric field on the sides of graphene exceeds the critical value.

It may be possible to stabilize the electrons on graphene by using fast oscillations as it was first proposed by P.L. Kapitsa for the pendulum with vibrating axis of suspension [17]. It turns out that the pendulum can perform stable vibrations around the point where it points upwards, which is unstable in the absence of the vibrations of the point of suspension. It was generalized to arbitrary oscillating potentials in [18]. The trapping of a classical particle can be achieved by introducing an external rapid time periodic potential [17,18] as follows:

$$V(r,t) = V_0(r) + V_1(r,t), \qquad (1)$$

where $r$ is the space coordinate. Hereafter, we consider electron trapping in $y$ direction around $y = 0$ (see Fig.1(C)). The classical particle is trapped by an effective time-independent potential which is approximately given by[17]

$$V_{eff}(y) = V_0(y) + \overline{F^2}(y)/(2m\omega^2), \qquad (2)$$

Where $\overline{F^2}(y)$ is the time average of the square of the force $F(y) = -\partial V_1(y,t)/\partial t$, which is exerted by the oscillating field $V_1(y,t) = V_1(y,t+T)$, where $T = 2\pi/\omega$, and its time average $\overline{V_1}$ vanishes. In this case, the trapping is obtained when the frequency of the external oscillatory field $\omega$ is much larger than the frequency $\Omega$ of the bound motion or the inverse of the shortest characteristic time scale of the motion, which for this comparison plays the role of $\Omega$ [17].

We consider an AC electric current in the graphene layer as a stabilizing fast oscillation. In Fig.2(A), it is shown a circuit where the graphene layer is incorporated into the LC circuit with the external driving AC source. This source produces an in-plane electric field $E_z(\omega)$ in the graphene layer which results in the periodic electron motion. The velocity of the electron can be taken as $v_z(\omega) = \mu_e E_z(\omega)$, where $\mu_e$ is the electron mobility in graphene. Hereafter, we consider a static approximation, where $\omega$ is much less than the plasma frequency of graphene. The collective motion of electrons in the z-direction produces a magnetic field $B = \mu_0 H$ around the graphene layer as illustrated in Fig.2(B). The magnetic flux density on the sides of graphene are given by

$$B_x = \frac{1}{2}\mu_0 J_s, \tag{3}$$

Where $J_s$ is the linear current density [A/m]. The magnetic field oscillates in time with the same frequency as the applied electric field. There is Lorentz force acting on the moving electrons in graphene given as follows:

$$F_y(\omega) = e \cdot v_z(\omega) \cdot B_x(\omega). \tag{4}$$

The direction of Lorentz force is towards the center of the layer regardless of the side of the graphene layer as illustrated in Fig.2(B). It is schematically shown the projection of the graphene layer in the X-Z plane. As an example, there is shown a moment of time where the electric current in the graphene layer is directed in the positive Z direction. The collective motion of electrons produces magnetic field which direction is along the positive X direction on the left side of the layer, and along the negative X direction on the right side of the layer. Thus, the direction of Lorentz force is towards the center of the layer on both sides. It should be also noted that Eq. (4) is valid /derived for electrically neutral current-carrying conductor [19].

Lorentz force is the restoring force to keep electrons localized on the graphene plate. It periodically squeezes the electrons to graphene and creates an effective time-independent potential according to Eq.(2). In Fig.3, we present the results of numerical modeling showing the trajectories of electrons in the presence of constant electric field $E_y$ and the time-varying magnetic field $B_x$. The trajectories are shown in Y-Z plane in normalized units $l_0 = m_e \cdot |v_{z,max}|/(e \cdot |B_{x,max}|)$, where $m_e$ is electron mass, $v_{z,max}$ is the maximum electron velocity in the Z direction, $e$ is the electron charge, $B_{x,max}$ is the maximum magnetic field in the X direction. In Fig.3(A), the black, the blue, the green, and the red curves correspond to the frequency of oscillation 1.0 $\omega_0$, 0.5 $\omega_0$, 0.2 $\omega_0$, and 0.1 $\omega_0$, respectively, where $\omega_0$ is the cyclotron frequency $\omega_0 = e \cdot |B_{x,max}|/m_e$. The strength of the electric field intensity $E_y$ is taken $E_y = 0.1 \cdot v_{z,max} B_{x,max}$ in all cases. The Y component of the trajectories increases with the decrease of the current frequency $\omega$. In Fig. 3(B), the blue, the black, and the red curves are obtained for the three levels of electric field intensity $E_y$ : 0.1 $E_0$, 0.5 $E_0$, and 1.0 $E_0$, where $E_0 = e \ |v_{z,max}| \cdot |B_{x,max}|$. The results in Fig.3(B) are obtained for $\omega = 1.0 \ \omega_0$. The increase of the magnitude of the electric field without the increase in the frequency of oscillation leads to the electrons escape from the middle plate.

Finally, we estimate the practically achievable value of the effective time-independent potential given by

$$V_{eff}(y) = \overline{F^2}(y)/(2m_e\omega^2). \qquad (5)$$

The linear current density is estimated based on the data reported in Ref. [20]. The maximum current density of $1.18 \times 10^8$ A/cm² was observed for 0.3 µm graphene interconnect on SiO₂/Si substrate. It leads to $J_s = 0.35 \times 10^5$ A/m linear current density and 44 mT maximum magnetic field around the layer. The maximum electron velocity is taken $|v_{z,max}| = 1 \times 10^7$ m/s based on the experimental data presented in Ref. [21]. The maximum value of the Lorentz force according to Eq.(4) is about 70 fN. The frequency of oscillations is taken to be $10\,\omega_0 = 7.7$ GHz. The effective potential given by Eq.(5) is $V_{eff}(y) \approx 1 \times 10^{-17}\,J$ or $60\,eV$. It is much larger than 25 meV corresponding to $k_b T$ at room temperature, where $k_b$ is the Boltzmann's constant. We want to emphasize that the estimated effective potential produced by the fast oscillations is not aimed at compensating the external force produced by the electric field $E_y$ but to make a stable state for electrons on the middle plate of the capacitor. In this case, the maximum strength of the electric field between the middle and the side plates of the three-plate capacitor as shown in Fig.1 (C), may exceed the breakdown value $E_b$ at normal conditions.

The increase in the maximum electric field intensity leads to an increase of the electric energy density $U_E = 1/2\,\epsilon_0 E^2$ stored between the plates of the capacitor. As long as the electrons on the middle plate are in the equilibrium state, the electric breakdown can be only due to the flow of positive ions from the side plates to the middle plate. The value of the ion electric breakdown can be estimated from the mechanical strength of the material as follows: $U_E \leq \sigma$, where $\sigma$ is the tensile strength of the material. It defines the amount of tensile stress a material can withstand before breaking. For example, taking the tensile strength of the side plate to be 300 MPa (e.g., the plates made of copper), the maximum electric energy density is 300 MJ/m³. This energy density corresponds to the electric field intensity of 8.2 GV/m. For comparison, the energy density of gasoline is 34.2 GJ/m³. Carbon nanotubes (CNTs) are known to be the strongest and stiffest materials with tensile strength in the range of 11-63 GPa [22]. Theoretically, the energy density in the triple plate capacitor with side plates made of metallic carbon nanotubes can exceed the one of gasoline. It would correspond to the electric field between the plates exceeding 100 GV/m.

The development of electric energy storage devices with energy density exceeding the one of gasoline would have a drastic impact on the development of human civilization. One can imagine all-electric aircraft and helicopters operating without burning gallons of gasoline, lifetime-charged smartphones and laptops. The ideas described in this work show just one of many possible approaches. The operation of the three-plate capacitor is based on the hypothesis of charge localization in the middle of the graphene plate (see Fig. 1(C)) and charge stabilization by AC electric current. The stabilization of electrons can be experimentally verified using the device schematically shown in Fig.4. It resembles graphene FET studied in a number of works [23]. The only modification is associated with the additional circuit to produce AC in the graphene layer. The effect of the pseudopotential produced by the oscillations should result in the decrease of the leakage graphene-to-gate current. The level of the leakage current suppression should depend on the amplitude as well as on the frequency of AC.

All the estimates in this work are based on the classical equations without accounting for the quantum mechanical phenomena. For instance, the quantum capacitance was predicted for a three-plate capacitor in which the middle plate represents a two-dimensional metal [16]. It may significantly change the total capacitance of the device. It should be also noted that Eq. (4) is valid /derived for electrically neutral current-carrying conductor [19]. These and many other questions deserve a special study. The main objective of this work is to consider the intriguing possibility of utilizing graphene properties for high-density electric energy storage.

To conclude, we considered a three-plate capacitor with a middle plate made of single-layer graphene. We hypothesize that electrons on the middle plate may experience a zero net force due to the unique property of graphene not to screen an external electrostatic field. We considered fast current oscillations to make electrons on the middle plate in a stable equilibrium state. AC electric current through the graphene layer produces a magnetic field. The arising Lorentz force squeezes the moving electrons to the graphene plate. The value of the pseudopotential may exceed $60\ eV$ due to the high mobility and large current density of graphene. Theoretically, our approach may lead to the enhancement of the electric energy density in electrostatic capacitors.

The data that support the findings of this study are available from the corresponding author upon reasonable request.

**Figure Captions**

Figure 1. (A) Schematics of the three-plate parallel plate capacitor where the central plate is made of a regular (bulk) metal. The plates are shown in X-Z plane. The side plates are charged positively while the middle plate is charged negatively. (B) The electron potential energy U along the Y-axis. (C) Schematics of the three-plate parallel plate capacitor where the central plate is made of a single-layer graphene. (D)

The electron potential energy U along the Y-axis. The picture of graphene layer is from https://commons.wikimedia.org/wiki/File:Graphen.jpg, contribution of AlexanderAlUS.

Figure 2. (A) Schematics of the three-plate capacitor where the middle plate made of graphene is incorporated into the LC circuit with the external driving AC source. This source produces A+C current in the graphene layer. (B) Schematics showing the direction of Lorentz force acting on the moving electrons. The direction of Lorentz force is towards the center of the layer on both sides of graphene layer.

Figure 3. The results of numerical modeling showing the trajectories of electrons in the presence of constant electric field $E_y$ and the time-varying magnetic field $B_x$. The trajectories are shown in Y-Z plane in normalized units $l_0$. (A) The black, the blue, the green, and the red curves correspond to the frequency of oscillation 1.0 $\omega_0$, 0.5 $\omega_0$, 0.2 $\omega_0$, and 0.1 $\omega_0$, where $\omega_0$ is the cyclotron frequency. (B) The blue, the black, and the red curves are obtained for the three levels of electric field intensity $E_y$ : 0.1 $E_0$, 0.5 $E_0$, and 1.0 $E_0$, where $E_0 = e \, |v_{z,max}| \cdot |B_{x,max}|$.

Figure 4. Schematics of the device for the proof-of-the-concept experiments. There is shown a graphene field effect transistor, where the graphene channel is included into the oscillatory circuit. It is expected that AC current in graphene will suppress the leakage channel-to-gate current.

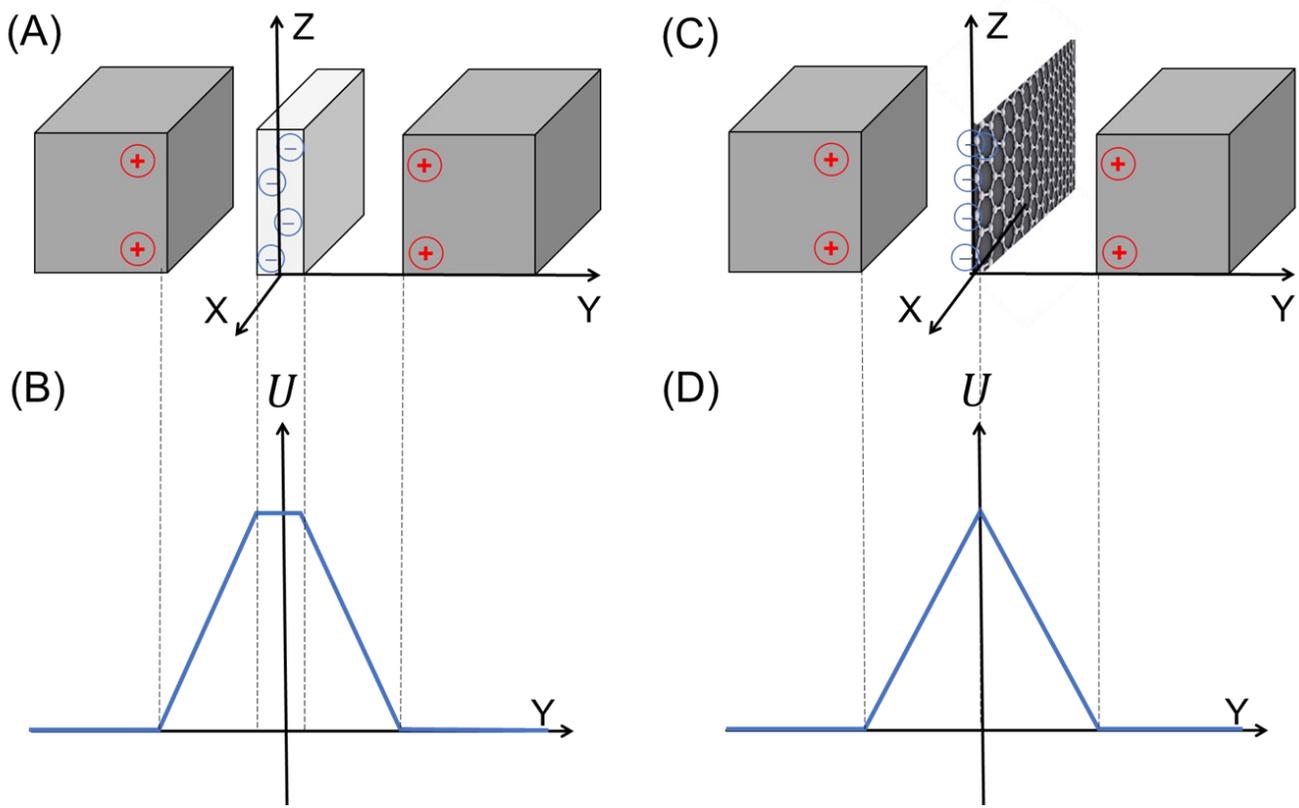

Figure 1

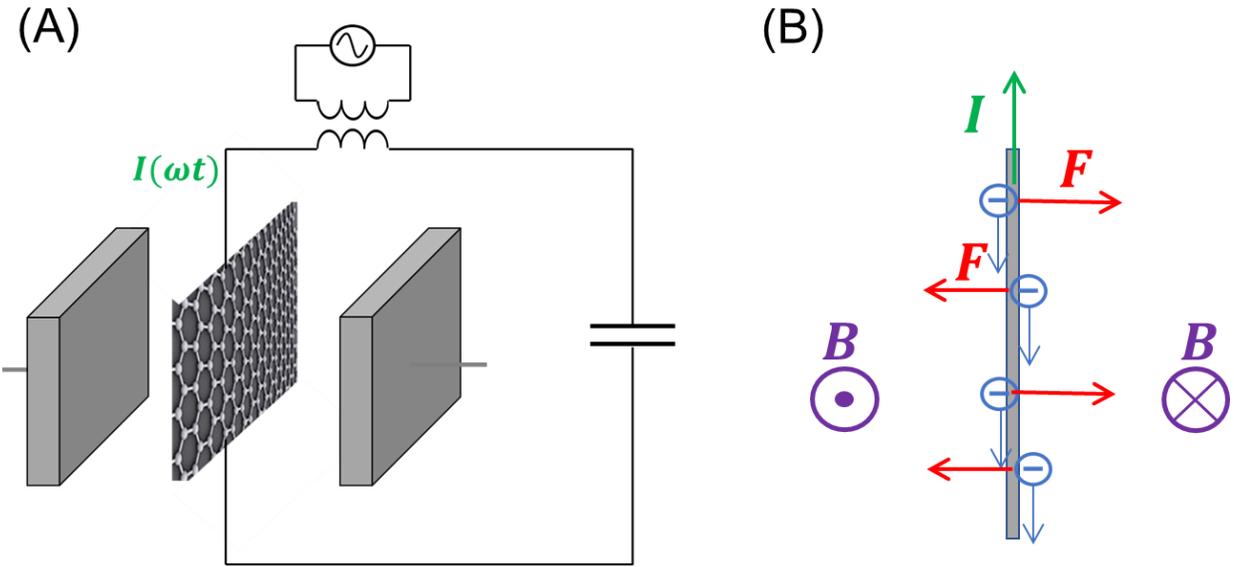

Figure 2

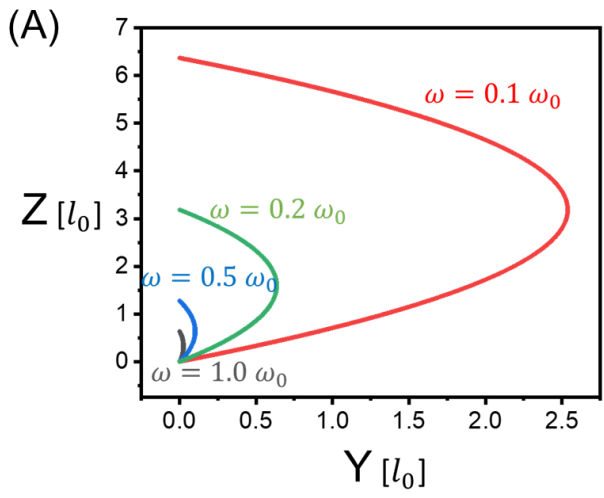 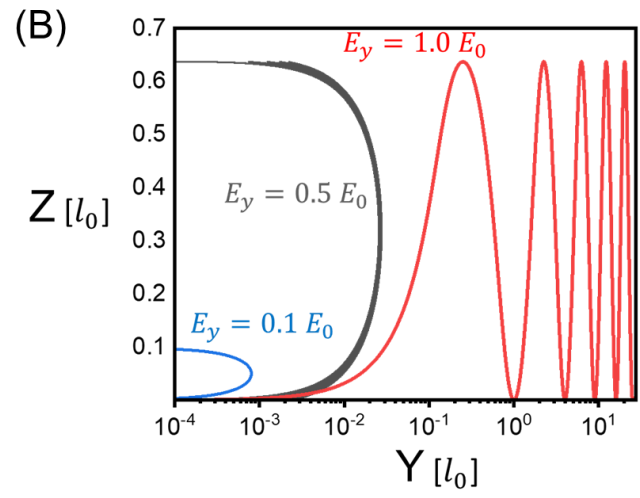

Figure 3

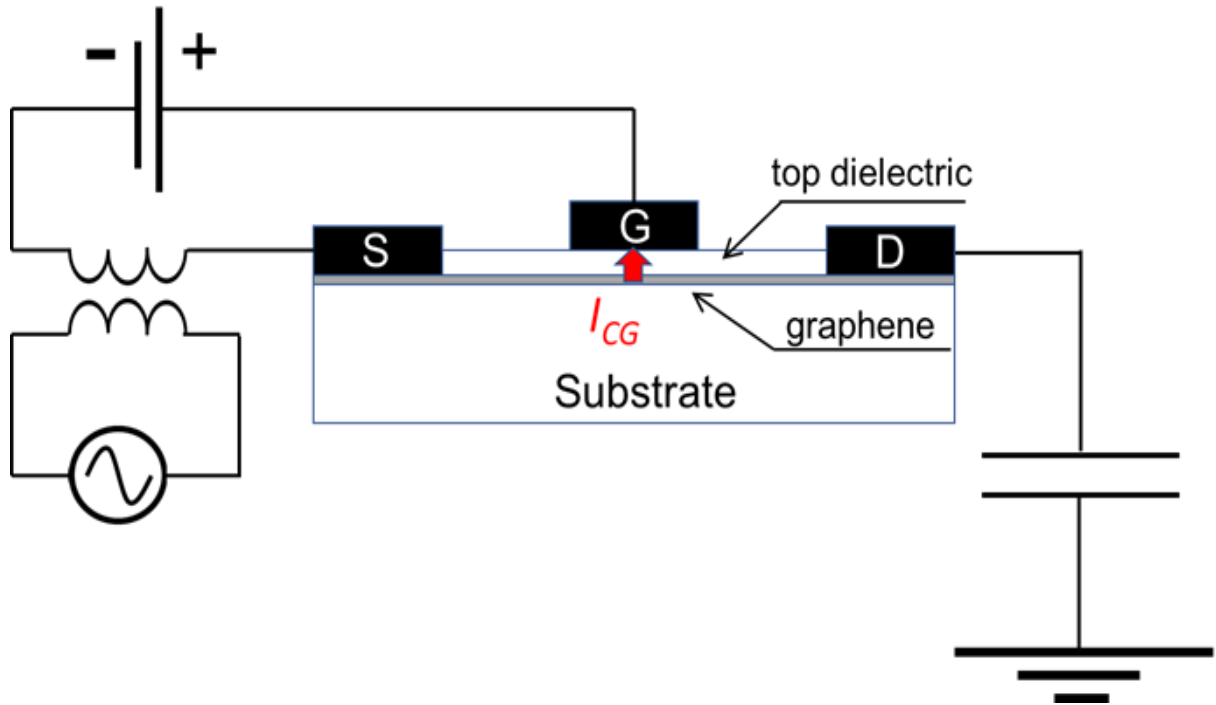

Figure 4